\documentclass{elsart}

\newcommand{ \be	}{\begin{equation}}
\newcommand{ \ee 	}{\end{equation}}
\newcommand{ \bea 	}{\begin{eqnarray}}
\newcommand{ \eea	}{\end{eqnarray}}
\newcommand{ \ba	}{\begin{array}}
\newcommand{ \ea 	}{\end{array}}
\newcommand{ \Ntrig	}{N_{\rm trig}}
\newcommand{ \Nmb	}{N_{\rm inc}}
\newcommand{ \Bmb	}{B_{\rm inc}}
\newcommand{ \Njet	}{N_{\rm jet}}
\newcommand{ \Nass	}{N_{\rm assoc}}
\newcommand{ \Nbg	}{N_{\rm bkgd}}
\newcommand{ \dphi	}{\Delta\phi}
\newcommand{ \phitrig	}{\phi_{\rm t}}
\newcommand{ \vtrig	}{v_2^{\rm (t)}}
\newcommand{ \vvtrig	}{v_4^{\rm (t)}}
\newcommand{ \vone	}{v_2^{\rm (1)}}
\newcommand{ \vvone	}{v_4^{\rm (1)}}
\newcommand{ \vtwo	}{v_2^{\rm (2)}}
\newcommand{ \vvtwo	}{v_4^{\rm (2)}}
\newcommand{ \vjet	}{v_2^{\rm (ta)}}
\newcommand{ \vvjet	}{v_4^{\rm (ta)}}
\newcommand{ \vnjet	}{v_2^{\rm (to)}}

\newcommand{ \pt	}{p_{\perp}}
\newcommand{ \mean	}[1]{\langle {#1} \rangle}
\newcommand{ \Jhat	}{\hat{Y}}

\usepackage{psfig}

\begin{document}

\begin{frontmatter}

\title{Analysis Method for Jet-Like Three-Particle Azimuthal Correlations}

\author{Jason G. Ulery}
\ead{ulery@physics.purdue.edu}
\author{Fuqiang Wang}
\ead{fqwang@purdue.edu}

\address{Department of Physics, Purdue University, 525 Northwestern Avenue, West Lafayette, Indiana 47907, USA}

\begin{abstract}
Jet-like three-particle azimuthal correlations can discriminate various physical scenarios that have been proposed to explain the observed strong modification to two-particle azimuthal correlations. The three-particle correlation analysis is notoriously difficult in heavy-ion collisions due to the large combinatorial backgrounds. We describe the general idea behind the jet-like three-particle azimuthal correlation analysis, with emphasis on the subtraction of the combinatorial backgrounds. We discuss in detail various sources of systematic uncertainties in such an analysis. 
\end{abstract}

\begin{keyword}
heavy-ion; jet; azimuthal correlation; three-particle

\PACS 25.75.-q, 25.75.Dw
\end{keyword}
\end{frontmatter}

\maketitle

\section{Introduction}

Two-particle azimuthal correlations, between one particle (trigger particle) at high transverse momentum ($\pt$) to trigger on a jet~\footnote{Note that we loosely refer to `jet' as particles clustered in azimuthal angle.} and the other (associated particle) at lower $\pt$, have become a powerful tool to study interactions between jets and the hot and dense medium created in relativistic heavy-ion collisions at RHIC. The measured two-particle azimuthal correlations were found to be significantly modified by the medium~\cite{b2b,jetspec,wangRIKEN,holtzman,lacey,wangMIT,uleryQM05,wangQM05,cole,jia}. This modification, in turn, can be used to study the properties of the medium. 

The away-side associated particles (opposite to the trigger particle in azimuth) in central Au+Au collisions, in particular, were found to be not narrowly distributed in azimuth at $\dphi=\pi$ from the trigger particle, as observed in $pp$ and d+Au collisions. Rather, the azimuthal distributions of the away-side associated particles are very broad, and perhaps exhibit a double-hump structure with a dip at $\dphi=\pi$~\cite{jetspec,wangRIKEN,holtzman,lacey,wangMIT,uleryQM05,wangQM05,cole,jia}. The away-side associated particles were also found to be significantly softened, not much harder than the bulk medium particles, indicating a partial thermalization between jet particles and the bulk medium~\cite{jetspec,wangRIKEN,holtzman,wangMIT,uleryQM05,wangQM05}. Moreover, the away-side associated particles were found to be softer at $\dphi=\pi$ than in the double-hump regions~\cite{wangRIKEN,wangMIT,uleryQM05,wangQM05}.

Several scenarios have been proposed to explain the observations. One scenario is deflection of jets by transverse radial flow of the bulk medium~\cite{armesto} or by the large survival probability of jet particles moving outwards than inwards due to the path-length dependence of energy loss~\cite{hwa}. Such a scenario would have jet particles narrowly clustered in individual events but the cluster is broadly distributed around $\dphi=\pi$ over many events. Another scenario is large angle gluon radiation~\cite{vitev}. Such a scenario would possess qualitatively similar structure as for deflected jets. A third scenario is conical flow generated by sound shock-waves due to large energy deposition by high momentum partons in the medium~\cite{stoecker,shuryak,muller}. Such shock-waves result in a distinctive Mach-cone type structure where particles are preferentially emitted at the Mach angle determined by the speed of sound in the medium, independent of the particle $\pt$. If Mach-cone type conical flow is indeed responsible for the observation, then the extraction of the speed of sound (thereby the equation of state) of the created medium may be possible. A fourth scenario is $\check{\rm C}$herenkov gluon radiation generated by interactions of fast particles with the medium~\cite{dremin,majumder}. Such a scenario would also result in conical emission of particles, as same as Mach-cone shock-waves, however, the $\check{\rm C}$herenkov angle would depend on the associated particle $\pt$~\cite{majumder}.

Two-particle correlations cannot distinguish these scenarios because they give, qualitatively, the same two-particle correlation function. Three-particle jet-like azimuthal correlations are needed to distinguish these scenarios by using the extra information offered by the third particle. If the broad two-particle correlation is due to deflected jets or large angle gluon radiation, the two associated particles will be narrowly clustered in angle but the cluster will swing over a wide range in azimuth on the away-side. If Mach-cone shock-waves or $\check{\rm C}$herenkov gluon radiation is responsible for the broad two-particle correlation, then particles will be emitted conically, thus the two associated particles will have equal probabilities being opposite away from $\dphi=\pi$ and being clustered together. A three-particle jet-correlation signal with opposite azimuthal angles from $\dphi=\pi$ for the two associated particles is therefore a distinctive signature of Mach-cone shock-waves or $\check{\rm C}$herenkov gluon radiation. The $\pt$ dependence of the cone angle will further discriminate between the two scenarios~\cite{majumder}.

In central heavy-ion collisions, jet-like correlations are overwhelmed by large combinatorial backgrounds due to the large particle multiplicity. The signal to background ratio depends on the collision centrality and the $\pt$ range of the associated particles; the typical signal to background ratio for a modest associated $\pt$ range of 1-2~GeV/$c$ is of the order of 1/30 for central Au+Au collisions. The three-particle correlation background will be much larger because the background level is squared; with the same associated $\pt$ range the three-particle signal to background ratio is of the order of $10^{-3}$ in central Au+Au collisions. Thus the three-particle correlation analysis is extremely challenging -- one needs a statistical and systematic precision of the order of $10^{-4}$; the precise subtraction of background is critical. In this paper, we describe an analysis method of three-particle jet-like correlations~\cite{uleryQM05,wangQM05} that is feasible to achieve such a precision.

The paper is organized as follows. Section~\ref{sec:method} describes the general ideas of the analysis method and background subtraction technique. Section~\ref{sec:realData} discusses complications in a real data analysis and the handling of backgrounds. Section~\ref{sec:systError} discusses the various sources of systematic uncertainties involved in such a data analysis and demonstrates that the systematics are well controlled with the background subtraction technique. Section~\ref{sec:summary} summarizes the paper.

\section{Analysis Method\label{sec:method}}

We suppose an event is composed, besides the trigger particle, of two components, the signal that is correlated with the trigger (such as a di-jet that is selected by the trigger particle but excluding the trigger particle itself) and the underlying background from the bulk medium (anything uncorrelated to the trigger particle except correlations due to anisotropic flow). Suppose the number of jet-correlated particles is $\Nass$, the number of uncorrelated background particles is $\Nbg$, and the total number of particles is $N=\Nass+\Nbg$. The total number of particle pairs is $N(N-1)$, which is composed of three parts: the number of background pairs $\Nbg(\Nbg-1)$, the number of jet-correlated pairs $\Nass(\Nass-1)$, and the number of jet particle and background particle pairs (cross pairs) $\Nass\Nbg+\Nbg\Nass=2\Nass\Nbg$. To obtain the number of jet-correlated pairs, one may subtract the numbers of background pairs and cross pairs from the total number of pairs~\footnote{We have used fixed multiplicities of $\Nass$ and $\Nbg$ (and $N=\Nass+\Nbg$) in the text, but the argument is also valid for average multiplicities except that the numbers of pairs are $\mean{\Nass}^2$, $\mean{\Nbg}^2$, and $\mean{N}^2$ for Poisson or Gaussian multiplicity distributions.}. Of course, if one is only interested in the number of jet-correlated pairs, one can easily get that from the number of jet-correlated particles, $\Nass$. Since we are interested in the angular correlation of the jet-correlated pairs that is unknown a priori, the only way to do is to subtract the angular correlations of the background pairs and the cross pairs. The angular correlations of the background pairs and the cross pairs can be obtained because they, unlike those of jet-correlated pairs, are not intrinsically correlated (except for the anisotropic flow correlation). Therefore, by subtracting from the raw three-particle correlation the angular correlations of the background pairs and the cross pairs, one obtains the genuine three-particle jet-correlation. 

Below we are restating the above with a set of equations. The two-particle jet-correlation is
\bea
\Jhat_2(\dphi)&=&Y_2(\dphi)-B_2(\dphi);\label{eq:2p1}\\
Y_2(\dphi)&=&dN/d\dphi,\label{eq:2p2}\\
B_2(\dphi)&=&d\Nbg/d\dphi,\label{eq:2p3} \\
\Jhat_2(\dphi)&=&d\Njet/d\dphi,\label{eq:2p4} 
\eea
where $\dphi=\phi-\phitrig$ is the associated particle azimuthal angle ($\phi$) relative to that of the trigger particle ($\phitrig$). We have used $Y_2$, $B_2$, and $\Jhat_2$ as shorthand notations for the raw particle azimuthal distribution (relative to the trigger particle azimuth) for the entire event (i.e., raw two-particle azimuthal correlation per trigger particle), the azimuthal distribution of the underlying background particles, and the azimuthal distribution of the associated jet particles (i.e., the two-particle jet-correlation per trigger particle), respectively.

The three-particle jet-correlation that we point out is the correlation between the trigger particle and two associated jet particles. What we have is the raw three-particle azimuthal correlation function,
\be
Y_3(\dphi_1,\dphi_2)=Y_2(\dphi_1)\otimes Y_2(\dphi_2)=\left[\Jhat_2(\dphi_1)+B_2(\dphi_1)\right]\otimes\left[\Jhat_2(\dphi_2)+B_2(\dphi_2)\right],\label{eq:J3}
\ee
where $\dphi_1=\phi_1-\phitrig$ and $\dphi_2=\phi_2-\phitrig$ are the azimuthal angles of the two associated particles ($\phi_1$ and $\phi_2$, respectively) relative to that of the trigger particle. We have used the notation of $A\otimes B$ to indicate correlation between A and B, not a simple product. The raw three-particle correlation function contains three pieces, the genuine three-particle jet-correlation that we discuss,
\be\Jhat_3(\dphi_1,\dphi_2)=\Jhat_2(\dphi_1)\otimes\Jhat_2(\dphi_2),\label{eq:J3hat}\ee
and two background terms. One is the combinatorial background between an associated jet particle and a background particle, 
\be
\Jhat_2\otimes B_2=\Jhat_2(\dphi_1)\otimes B_2(\dphi_2)+B_2(\dphi_1)\otimes\Jhat_2(\dphi_2),\label{eq:J2hatXB2}
\ee
which we term as the {\em hard-soft} background. 
If the jet-correlated particle and the background particle are uncorrelated, then we can remove the symbol $\otimes$ from Eq.(\ref{eq:J2hatXB2}), and readily obtain the hard-soft background from the simple product of the two-particle jet-correlation function $\Jhat_2$ and the background $B_2$ as
\be
\Jhat_2\otimes B_2=\Jhat_2(\dphi_1)B_2(\dphi_2)+B_2(\dphi_1)\Jhat_2(\dphi_2).\label{eq:J2hatB2}
\ee
However, induced correlations between jet-correlated particles and background particles can be expected, for instance, due to their correlations both to the orientation of the trigger particle relative to the reaction plane angle and to the event centrality (impact parameter). Thus in principle the hard-soft background can only be obtained by
\be
\Jhat_2\otimes B_2=\int_{0}^{\infty}\frac{1}{\sigma_{\rm tot}}\frac{d\sigma}{db}db\int_{0}^{2\pi}\frac{d\Psi}{2\pi}\int_{0}^{2\pi}\frac{d\phitrig}{2\pi}\left[
\ba{l}
\Jhat_2(\dphi_1;\phitrig,\Psi,b,...)B_2(\dphi_2;\phitrig,\Psi,b,...)+\nonumber\\
B_2(\dphi_1;\phitrig,\Psi,b,...)\Jhat_2(\dphi_2;\phitrig,\Psi,b,...)\nonumber
\ea
\right],\label{eq:J2hatXB2_exact}
\ee
where $d\sigma$ and $\sigma_{\rm tot}$ are the differential and total cross-sections, respectively, $b$ is the impact parameter, and $\Psi$ is the reaction plane angle (impact parameter direction). There could be other event characteristic variables besides $b$ and $\Psi$. The handling and effects of those induced correlations will be discussed later in Secions~\ref{sec:jetAnisotropy} and~\ref{sec:inducedCorrel}, respectively. For the sake of easy description and relatively simple equations, we will assume, for now, no correlation between jet-correlated particles and background particles, and carry on our study using Eq.(\ref{eq:J2hatB2}). 

The other background term is the combinatorial background between two background particles,
\be B_3(\dphi_1,\dphi_2)=B_2(\dphi_1)\otimes B_2(\dphi_2),\label{eq:B3B2B2}\ee
which we term as the {\em soft-soft} background. The soft-soft background can be obtained from inclusive events as we will discuss in Section~\ref{sec:realData.C}. Once the two background terms are determined, the final three-particle correlation is simply obtained by
\be
\Jhat_3(\dphi_1,\dphi_2)=Y_3(\dphi_1,\dphi_2)-\left[\Jhat_2(\dphi_1)B_2(\dphi_2)+B_2(\dphi_1)\Jhat_2(\dphi_2)\right]-B_3(\dphi_1,\dphi_2).\label{eq:J3hatBg}
\ee

\subsection{Elliptic Flow Correction}

In heavy-ion collisions, the background particles are correlated to the reaction plane due to the hydrodynamic kind of collective flow of the bulk medium and the anisotropic overlap region between the colliding nuclei. The trigger particle emission is also correlated to the reaction plane due to, not so much of the hydrodynamic collective flow, but the path-length dependent energy loss of high $\pt$ particles in the medium that is initially anisotropic. This reaction plane correlation at mid-rapidity, expressed in harmonics up to the fourth order, is given by
\be\frac{d\Nbg}{d\dphi}=\frac{\Nbg}{2\pi}[1+2v_2\cos2(\phi-\Psi)+2v_4\cos4(\phi-\Psi)],\label{eq:flow2p}\ee
where $\phi$ is the azimuthal angle of the particle and $\Psi$ that of the reaction plane. The reaction plane distribution is uniform. The first and the third harmonic terms are, due to symmetry, zero at mid-rapidity for symmetric collision systems which we consider in the present work. 

The distribution of the number of trigger-background particle pairs is given by
\bea
\frac{d^4N}{d\Psi d\phitrig d\phi d\dphi}=\frac{1}{2\pi}&\cdot&\frac{\Ntrig}{2\pi}\left[1+2\vtrig\cos2(\phitrig-\Psi)+4\vvtrig\cos4(\phitrig-\Psi)\right]\nonumber\\
&\cdot&\frac{\Nbg}{2\pi}\left[1+2v_2\cos2(\phi-\Psi)+4v_4\cos4(\phi-\Psi)\right]\cdot\delta(\dphi-\phi+\phitrig)\label{eq:flow2p1}
\eea
for a given set of the reaction plane angle $\Psi$, the trigger azimuthal angle $\phitrig$, and the associated particle azimuthal angle $\phi$. Here $\Ntrig$ is the total number of trigger particles, $\Nbg$ is the number of background particles per event, $v_2$ and $v_4$ are the second (elliptic flow) and fourth harmonic coefficients of the background particles, and the superscript `(t)' denotes those of the trigger particles. Integrating over $\Psi$, $\phitrig$, $\phi$, we obtain
\bea
\frac{dN}{d\dphi}&=&\int_0^{2\pi}d\Psi\int_0^{2\pi}d\phi\int_0^{2\pi}d\phitrig\frac{d^4N}{d\Psi d\phitrig d\phi d\dphi}\nonumber\\
&=&\frac{\Ntrig\Nbg}{2\pi}\left[1+2\vtrig v_2\cos(2\dphi)+2\vvtrig v_4\cos(4\dphi)\right].\label{eq:flow2p2}
\eea
Normalized per trigger particle, the anisotropic flow modulation is therefore
\be B_2(\dphi)=B_1\left[1+2\vtrig v_2\cos(2\dphi)+2\vvtrig v_4\cos(4\dphi)\right],\label{eq:bg2p}\ee
where $B_1=\Nbg/2\pi$ is the background azimuthal density.

The distribution of trigger-background-background triplets is given by
\bea
\frac{d^6N}{d\Psi d\phitrig d\phi_1d\phi_2d\dphi_1d\dphi_2}=\frac{1}{2\pi}&\cdot&\frac{\Ntrig}{2\pi}\left[1+2\vtrig\cos2(\phitrig-\Psi)+4\vvtrig\cos4(\phitrig-\Psi)\right]\nonumber\\
&\cdot&\frac{N_1}{2\pi}\left[1+2\vone\cos2(\phi_1-\Psi)+4\vvone\cos4(\phi_1-\Psi)\right]\label{eq:flow3p1}\\
&\cdot&\frac{N_2}{2\pi}\left[1+2\vtwo\cos2(\phi_2-\Psi)+4\vvtwo\cos4(\phi_2-\Psi)\right]\nonumber\\
&\cdot&\delta(\dphi_1-\phi_1+\phitrig)\cdot\delta(\dphi_2-\phi_2+\phitrig).\nonumber
\eea
for a given set of $\Psi$, $\phitrig$, $\phi_1$, and $\phi_2$. Integrating over $\Psi$, $\phitrig$, $\phi_1$, and $\phi_2$, we obtain
\bea
\frac{d^2N}{d\dphi_1 d\dphi_2}&=&\int_0^{2\pi}d\Psi\int_0^{2\pi}d\phitrig\int_0^{2\pi}d\phi_1\int_0^{2\pi}d\phi_2\frac{d^6N}{d\Psi d\phitrig d\phi_1 d\phi_2 d\dphi_1d\dphi_2}=\frac{\Ntrig N_1 N_2}{(2\pi)^2}\times\nonumber\\
&&\hspace{-0.2in}\left(
\ba{ll}
1&+2\vtrig\vone\cos(2\dphi_1)+2\vtrig\vtwo\cos(2\dphi_2)+2\vone\vtwo\cos2(\dphi_1-\dphi_2)\\
 &+2\vvtrig\vvone\cos(4\dphi_1)+2\vvtrig\vvtwo\cos(4\dphi_2)+2\vvone\vvtwo\cos4(\dphi_1-\dphi_2)\\
 &+2\vtrig\vone\vvtwo\cos2(\dphi_1-2\dphi_2)+2\vtrig\vtwo\vvone\cos2(2\dphi_1-\dphi_2)\\
 &+2\vone\vtwo\vvtrig\cos2(\dphi_1+\dphi_2)
\ea
\right).\label{eq:flow3p2}
\eea
Here we have taken the two background particles to be from two separate sets, with $N_1$ and $N_2$ the respective multiplicities. The product $N_1 N_2$ is the number of pairs. For particles from the same set, this becomes $\mean{\Nbg(\Nbg-1)}$. Normalized per trigger particle, the anisotropic flow modulation is therefore,
\bea
B_3(\dphi_1,\dphi_2)&=&\frac{\mean{\Nbg(\Nbg-1)}}{(2\pi)^2}\times\nonumber\\
&&\hspace{-0.4in}\left(
\ba{ll}
1&+2\vtrig\vone\cos(2\dphi_1)+2\vtrig\vtwo\cos(2\dphi_2)+2\vone\vtwo\cos2(\dphi_1-\dphi_2)\\
 &+2\vvtrig\vvone\cos(4\dphi_1)+2\vvtrig\vvtwo\cos(4\dphi_2)+2\vvone\vvtwo\cos4(\dphi_1-\dphi_2)\\
 &+2\vtrig\vone\vvtwo\cos2(\dphi_1-2\dphi_2)+2\vtrig\vtwo\vvone\cos2(2\dphi_1-\dphi_2)\\
 &+2\vone\vtwo\vvtrig\cos2(\dphi_1+\dphi_2)
\ea
\right).\label{eq:flow3p}
\eea

\subsection{Hard-Soft and Soft-Soft Backgrounds}

The two-particle jet correlation signal can be obtained from Eq.~(\ref{eq:2p1}) and (\ref{eq:bg2p}) as
\be
\Jhat_2(\dphi)=Y_2(\dphi)-B_1\left[1+2\vtrig v_2\cos(2\dphi)+2\vvtrig v_4\cos(4\dphi)\right].
\label{eq:J2hatBg}
\ee

The hard-soft background is given, as in Eq.(\ref{eq:J2hatB2}), by folding the two-particle jet-correlation signal of Eq.(\ref{eq:J2hatBg}) with the underlying background of Eq.(\ref{eq:bg2p}), namely
\be\ba{ll}
\Jhat_2\otimes B_2=&
\left(Y_2(\dphi_1)-B_1\left[1+2\vtrig\vone\cos(2\dphi_1)+2\vvtrig\vvone\cos(4\dphi_1)\right]\right)\times\nonumber\\
&\left(B_1\left[1+2\vtrig\vtwo\cos(2\dphi_2)+2\vvtrig\vvtwo\cos(4\dphi_2)\right]\right)+\nonumber\\
&\left(Y_2(\dphi_2)-B_1\left[1+2\vtrig\vtwo\cos(2\dphi_2)+2\vvtrig\vvtwo\cos(4\dphi_2)\right]\right)\times\nonumber\\
&\left(B_1\left[1+2\vtrig\vone\cos(2\dphi_1)+2\vvtrig\vvone\cos(4\dphi_1)\right]\right).\label{eq:HS}
\ea\ee

The soft-soft background is given by Eq.~(\ref{eq:flow3p}). Obviously the anisotropic flow is the only considered correlation in the soft-soft term of Eq.~(\ref{eq:flow3p}). There are other correlation effects in the real collision data, which we will discuss below in section~\ref{sec:realData.C}.

\section{Complications in Real Data Analysis\label{sec:realData}}

In real data analysis, the events are often binned in centrality. The centrality is often defined in terms of particle multiplicity. The background shape can be obtained from mixed-events, where the trigger particle is mixed with inclusive events~\footnote{The term ``inclusive events" refers to events from the minimum bias event sample within the same centrality class as those with a trigger particle but without specifically requiring a trigger in them.} from the same centrality bin; it can be also obtained from single particle distributions of minimum-bias events folded with a random trigger particle. These two should give the same result, however, the former may be preferred if subtle two-particle acceptance effects are present that cannot be taken care of in the single particle folding. 

Before event-mixing or particle folding, the particle azimuthal acceptance has to be corrected such that the azimuthal distribution average over many events is uniform. This correction should be applied to both trigger particles and associated particles. It is equivalent, in the case of two-particle correlation analysis, to the two-particle acceptance correction as is usually done in data analysis. For three-particle correlation, however, this cannot be simply corrected for by the folding of two two-particle acceptance corrections.

The analysis of three-particle correlation is extremely challenging because of the large combinatorial backgrounds. For central Au+Au collisions, the number of charged particles at mid-rapidity within $|\eta|<1$ for a modest $\pt$ range of $1<\pt<2$~GeV/$c$ is about 150; the di-jet charged hadron multiplicity within the same $\pt$ range for a trigger particle $\pt$ range of $3<\pt<4$~GeV/$c$ is of the order of a few ($\sim 5$)~\cite{uleryQM05,uleryHP}. In other words, the signal to background ratio is of the order of 1/30. Thus, the hard-soft background is a factor of 30 larger than the three-particle correlation strength, and the soft-soft background is a factor of $10^3$ larger. In order to have a significant three-particle correlation result, one needs a precision of $10^{-4}$, both statistically and systematically. Such a precision is unprecedented in relativistic heavy-ion collisions; the background subtraction is crucial in three-particle correlation analysis.

\subsection{Two-Particle Correlation Background}

In real data from relativistic heavy-ion collisions, because exclusive jet reconstruction is impossible at low jet energy, the background level and the true jet signal strength are unknown a priori. One has to make an {\em ad hoc} assumption about the background level and hence the signal strength. The often used assumption is that the jet-correlation signal is zero at some $\dphi$ region. The two-particle azimuthal correlation function measured at RHIC is the lowest around $\dphi=1$. The STAR experiment makes the assumption that the jet signal is zero within the fixed range of $0.8<|\dphi|<1.2$ (which we will refer to as zero-yield-at-1, or ZYA1)~\cite{jetspec}. The PHENIX experiment uses the so-called zero-yield-at-minimum (ZYAM) method in which the $\dphi$ region where the signal minimum resides is determined by the data itself~\cite{lacey}. 

There are many sources of background, including the hydrodynamic type underlying event consisting of mostly soft particles, as well as hard particles from other jets not selected by the trigger particle.
Since the underlying background event is uncorrelated with the di-jet selected by the trigger particle, the underlying background event should be represented by the inclusive events (with a caveat discussed below). This is not obvious at the first glance because, for example, the underlying event can contain other jets that are not selected by the trigger particle as aforementioned. But so do the inclusive events. We can assure ourselves by realizing the following. Suppose the number of jets (with a trigger particle in the selected $\pt$ range) is Poisson distributed with an average of $\mean{n}$ in inclusive events, the probability to have $n$ jets per event is $P_n=\mean{n}^n e^{-\mean{n}}/n!$. We select events with at least one jet (with a trigger particle in the selected $\pt$ range). The probability to have $n+1$ such jets (that also contain trigger particles) in the same event is $P_{n+1}=\mean{n}^{n+1} e^{-\mean{n}}/(n+1)!$. This is equivalently the probability to have a triggered event~\footnote{We use ``triggered event'' to refer to the event selected by a high $\pt$ trigger particle in off-line data analysis. We do not mean an event triggered on-line.} (with a trigger particle) with $n$ jets (that have trigger particles) in the underlying background event that is not associated with {\em the} selected trigger particle. But this event is going to be used $n+1$ times (because all trigger particles, not just the highest $\pt$ particle, are used in the analysis), so the probability to have $n$ jets in the underlying background event, now normalized by the total number of trigger particles, is given by $(n+1)P_{n+1}/\sum_{n=0}^{\infty}(n+1)P_{n+1}=\mean{n}^n e^{-\mean{n}}/n!$. This is identical to $P_n$, the probability to have $n$ jets per event for the inclusive events.

Since the inclusive events represent the underlying background, the background shape can be constructed by mixed-event technique, mixing trigger particles with all particles (within the associated particle $\pt$ range) from the inclusive events of the same centrality. Because the event plane correlation is lost in such event-mixing~\cite{noteRP}, the flow modulation needs to be added in by hand. 

There is one caveat, though, in using the inclusive events to represent the underlying background in triggered event. This is often referred to as ``trigger bias'': because of the centrality cuts often used in data analysis, and because the production of soft particles is proportional to the number of participants whilst that of high-$\pt$ particles is proportional to the number of binary collisions (but is strongly suppressed in central heavy-ion collisions), the inclusive events (that are from the same centrality class as those triggered events) do not exactly represent the multiplicity of the underlying background event. Thus, the background level needs to be scaled with the aforementioned working assumption of two-particle correlation ZYA1 (or ZYAM):
\be B_1=a\Bmb . \label{eq:B1} \ee
Here the scaling factor $a$ is the ratio of the underlying background multiplicity (in the triggered events) to the inclusive event multiplicity for a given centrality:
\be a=\Nbg/\Nmb .\label{eq:a} \ee

The background estimated from the two-particle ZYA1 (or ZYAM) method is likely an upper limit because the jet signal, due to its broadening, may be positive in all azimuth. As will be discussed in Section~\ref{sec:systError.5}, the three-particle correlation contains richer information because of the presence of a third particle. One may require the three-particle correlation to be positive definite and use three-particle correlation ZYAM to obtain the scaling factor $a$. The obtained $a$ will be smaller than that obtained from two-particle ZYAM, and may be more accurate in representing the underlying jet-correlation background.

\subsection{Hard-Soft Background in three-particle Correlation}

As discussed above we use $a\Bmb$ to estimate the multiplicity level $B_1$ of the underlying background in triggered events. The two-particle correlation is thus given by Eq.~(\ref{eq:J2hatBg}) with $B_1$ replaced by $a\Bmb$:
\be\Jhat_2(\dphi)=Y_2(\dphi)-a\Bmb\left[1+2\vtrig v_2\cos(2\dphi)+2\vvtrig v_4\cos(4\dphi)\right].\label{eq:J2hat}\ee
Likewise the hard-soft background is given by Eq.~(\ref{eq:HS}) with $B_1$ replaced by $a\Bmb$:
\be\ba{ll}
\Jhat_2\otimes B_2=&
\left(Y_2(\dphi_1)-a\Bmb\left[1+2\vtrig\vone\cos(2\dphi_1)+2\vvtrig\vvone\cos(4\dphi_1)\right]\right)\times\nonumber\\
&\left(a\Bmb\left[1+2\vtrig\vtwo\cos(2\dphi_2)+2\vvtrig\vvtwo\cos(4\dphi_2)\right]\right)+\nonumber\\
&\left(Y_2(\dphi_2)-a\Bmb\left[1+2\vtrig\vtwo\cos(2\dphi_2)+2\vvtrig\vvtwo\cos(4\dphi_2)\right]\right)\times\nonumber\\
&\left(a\Bmb\left[1+2\vtrig\vone\cos(2\dphi_1)+2\vvtrig\vvone\cos(4\dphi_1)\right]\right).\label{eq:J2B2}
\ea\ee

In real data analysis, the background is often obtained from mixed-events, mixing trigger particles with inclusive events as mentioned above. The event-mixing technique is used to account for any detector effects, which are present in the correlation signal. The flow modulation is often added dynamically pair-by-pair to avoid potential systematic effect of $\mean{\vtrig v_2}\neq\mean{\vtrig}\mean{v_2}$. However, such a systematic effect is found to be very small in real data.

\subsection{Soft-Soft Background in three-particle Correlation\label{sec:realData.C}}

The soft-soft background can be obtained by mixing inclusive events with a random trigger (from a triggered event), expressed in two $\phi$ angles relative to the random trigger particle:
\be
B_3^{\rm inc}(\dphi_1,\dphi_2)=\frac{1}{\Ntrig}\int_0^{2\pi}d\phitrig\frac{d^3\Nmb(\phitrig,\dphi_1,\dphi_2)}{d\phitrig d\dphi_1d\dphi_2}.\label{eq:B3inc}
\ee
This is effectively the two-particle distribution $B_3^{\rm inc}(\phi_1,\phi_2)$ in the two azimuthal angles but with the origin of the azimuthal angles shifted to the random trigger particle direction. This event-mixing procedure takes into account subtle acceptance effects that may not be fully corrected for by the single particle acceptance corrections. If the acceptance corrections are perfect so that the corrected single particle distributions of trigger and associated particles are perfectly uniform in azimuth, then the event-mixing is unnecessary and $B_3^{\rm inc}(\dphi_1,\dphi_2)\equiv B_3^{\rm inc}(\phi_1,\phi_2)$.

In real events, the background particles contain not only anisotropic flow correlation, but also other correlations that are unrelated to the trigger particle, such as jet-correlations due to jets other than the di-jet selected by the trigger particle~\cite{minijet}. Those correlations (other than the flow correlation) are contained in the inclusive events used in the event-mixing, therefore are already included in the soft-soft background in Eq.~(\ref{eq:B3inc}).

The anisotropic flow correlation between the two soft particles are also already contained in the inclusive events being mixed, but the flow correlation between the soft particles and the trigger particle is lost because of the randomization of the reaction planes in the mixing~\cite{noteRP}. The trigger-flow correlation needs to be added in by hand for each trigger-background-backgrounbd triplet in the event-mixing via Eq.~(\ref{eq:flow3p}) except the two terms that do not contain the trigger particle flow. The trigger-flow modulation that needs to be added in by hand is given by:
\bea
B_3^{\rm inc,tf}&=&\frac{\mean{\Nmb(\Nmb-1)}}{(2\pi)^2}\times\\
&&\hspace{-0.3in}\left(
\ba{l}
2\vtrig\vone\cos(2\dphi_1)+2\vtrig\vtwo\cos(2\dphi_2)+2\vvtrig\vvone\cos(4\dphi_1)+2\vvtrig\vvtwo\cos(4\dphi_2)+\nonumber\\
2\vtrig\vone\vvtwo\cos2(\dphi_1-2\dphi_2)+2\vtrig\vtwo\vvone\cos2(2\dphi_1-\dphi_2)+\\
2\vone\vtwo\vvtrig\cos2(\dphi_1+\dphi_2)
\ea
\right).\label{eq:B3inctf}
\eea
The trigger-flow correlation can be constructed by mixing the trigger particle with two inclusive, within the same centrality window but otherwise random, events.

The multiplicity of the inclusive events, $\mean{\Nmb}$, does not exactly represent the underlying background multiplicity of the triggered events, $\mean{\Nbg}$. Thus, we scale the soft-soft background constructed from the inclusive events, given by the sum of Eqs.~(\ref{eq:B3inc}) and (\ref{eq:B3inctf}), by the ratio of the number of pairs in the underlying background to that in the mixed inclusive event:
\be\frac{\mean{\Nbg(\Nbg-1)}}{\mean{\Nmb(\Nmb-1)}}\approx a^2,\label{eq:a2}\ee
where $a\approx 1$ is given by Eq.~(\ref{eq:a}). Eq.~(\ref{eq:a2}) would be exact if the multiplicity distributions are Poisson (or Gaussian in the large multiplicity limit) for which $\mean{N(N-1)}=\mean{N}^2$. Even if the multiplicity distributions are not Poisson, as long as $\frac{\mean{\Nbg(\Nbg-1)}}{\mean{\Nbg}^2}=\frac{\mean{\Nmb(\Nmb-1)}}{\mean{\Nmb}^2}$, Eq.(\ref{eq:a2}) would still be exact. However, in real data, the multiplicity distributions in the inclusive events and the underlying background of the triggered events can deviate differently from Poisson distributions. To take this into account, a second scaling factor $b$ is applied in front of $a^2$, namely
\be\frac{\mean{\Nbg(\Nbg-1)}}{\mean{\Nmb(\Nmb-1)}}= ba^2.\label{eq:ba2}\ee
Here the factor $b$ characterizes the effective difference in the associated multiplicity distributions of the inclusive events and the underlying background of triggered events:
\be b=\frac{\mean{\Nbg(\Nbg-1)}\left/\mean{\Nbg}^2\right.}{\mean{\Nmb(\Nmb-1)}\left/\mean{\Nmb}^2\right.}.\label{eq:b}\ee
The value of $b$ should be very close to 1. Due to the large combinatorics, even a small change in $b$ could have a significant effect on the magnitude of the final three-particle correlation result. However, the shape of the three-particle correlation is not much effected by small changes in $b$ because the soft-soft background is largely uniform. Since triggered events in heavy-ion collisions are dominated by the underlying background, we assume the characteristics of the associated multiplicity distributions in the entire triggered event and in the underlying background to be the same, $\frac{\mean{\Nbg(\Nbg-1)}}{\mean{\Nbg}^2}=\frac{\mean{N(N-1)}}{\mean{N}^2}$, and obtain $b$ as
\be b=\frac{\mean{N(N-1)}\left/\mean{N}^2\right.}{\mean{\Nmb(\Nmb-1)}\left/\mean{\Nmb}^2\right.}.\label{eq:btrig}\ee
Thus the complete soft-soft background is given by
\be B_3=ba^2\left[B_3^{\rm inc}(\dphi_1,\dphi_2)+B_3^{\rm inc,tf}(\dphi_1,\dphi_2)\right].\label{eq:B3}\ee

The trigger-flow by Eq.(\ref{eq:B3inctf}) would be exact if the background particles were only correlated by flow. However, additional correlations are present between the background particles in real data as already included in Eq.(\ref{eq:B3inc}), such as those due to resonance decays and background (mini-)jets. Those additional correlations alter the trigger-flow background. The effect can be corrected for to the first order by applying a multiplication factor $B_3^{\rm inc}/B_3^{\rm inc,sf}$ in front of $B_3^{\rm inc,tf}$ given by Eq.(\ref{eq:B3inctf}). Here $B_3^{\rm inc,sf}$ would be the soft-soft background in analogy to $B_3^{\rm inc}$ of Eq.(\ref{eq:B3inc}) if all correlations were due to flow. It can be obtained from the measured soft particle flow, keeping the number of soft pairs as same as in $B_3^{\rm inc}$, by
\bea
B_3^{\rm inc,sf}(\dphi_1,\dphi_2)&=&\frac{\mean{\Nmb(\Nmb-1)}}{(2\pi)^2}\times\nonumber\\
&&\left[
1+2\vone\vtwo\cos2(\dphi_1-\dphi_2)+2\vvone\vvtwo\cos4(\dphi_1-\dphi_2)\right].\label{eq:B3incflow}
\eea
Therefore, the complete soft-soft background in Eq.(\ref{eq:B3}) is modified as
\be B_3=ba^2 B_3^{\rm inc}(\dphi_1,\dphi_2)\left[1+\frac{B_3^{\rm inc,tf}(\dphi_1,\dphi_2)}{B_3^{\rm inc,sf}(\dphi_1,\dphi_2)}\right].\label{eq:B3new}\ee
And finally the three-particle correlation function is given by Eqs.~(\ref{eq:J3hatBg}), (\ref{eq:J2B2}), and (\ref{eq:B3new}).

\section{Sources of Systematics\label{sec:systError}}

In this section, we discuss the various sources of systematic uncertainties involved in the three-particle jet-like azimuthal correlation analysis. This is not meant to be a complete list of systematics, but we think these are the major ones. We conclude that the dominant systematic uncertainties come from uncertainties in the elliptic flow measurements and background normalization as discussed in sections~\ref{sec:systError.1} and \ref{sec:systError.5}, respectively.

\subsection{Elliptic Flow Uncertainties\label{sec:systError.1}}

The anisotropic flow correction has effect on the final three-particle azimuthal correlation function, because the number of extracted associated particles depends on the amount of flow subtraction~\cite{jetspec}. The size of the effect on the three-particle correlation function is similar to that on the two-particle correlation function. The sizes of the flow corrections in the two background terms (hard-soft and soft-soft) separately are, however, large compared to the magnitude of the genuine three-particle correlation, but the leading terms of the flow corrections in the hard-soft and soft-soft backgrounds cancel. We illustrate this by examining the total background, sum of Eq.~(\ref{eq:HS}) and (\ref{eq:flow3p}):
\bea
\Jhat_2\otimes B_2+B_3&=&Y_2(\dphi_1)\times B_1\left[1+2\vtrig\vtwo\cos(2\dphi_2)+2\vvtrig\vvtwo\cos(4\dphi_2)\right]+\nonumber\\
&&\hspace{-0.2in}Y_2(\dphi_2)\times B_1\left[1+2\vtrig\vone\cos(2\dphi_1)+2\vvtrig\vvone\cos(4\dphi_1)\right]-\nonumber\\
&&\hspace{-0.2in}2B_1^2\left[1+2\vtrig\vone\cos(2\dphi_1)+2\vvtrig\vvone\cos(4\dphi_1)\right]\times\nonumber\\
&&\hspace{0.1in}\left[1+2\vtrig\vtwo\cos(2\dphi_2)+2\vvtrig\vvtwo\cos(4\dphi_2)\right]+\nonumber\\
&&\hspace{-0.2in}B_1^2\left(
\ba{ll}
1&+2\vtrig\vone\cos(2\dphi_1)+2\vtrig\vtwo\cos(2\dphi_2)\\
 &+2\vvtrig\vvone\cos(4\dphi_1)+2\vvtrig\vvtwo\cos(4\dphi_2)\\
 &+2\vtrig\vone\vvtwo\cos2(\dphi_1-2\dphi_2)+2\vtrig\vtwo\vvone\cos2(2\dphi_1-\dphi_2)\\
 &+2\vone\vtwo\vvtrig\cos2(\dphi_1+\dphi_2)
\ea
\right).\label{eq:J2B2B3a}
\eea
Here we have dropped the anisotropic flow correlation between the two soft particles because they are already included in the inclusive events used in constructing the soft-soft background. In other words, the flow between the soft particles has been {\em precisely} included by the background inclusive event itself. We have also taken the number of background pairs per $(2\pi)^2$ to be $B_1^2$ for simple illustration. Rearranging the terms gives
\bea
\Jhat_2\otimes B_2+B_3&=&Y_2(\dphi_1)\times B_1\left[1+2\vtrig\vtwo\cos(2\dphi_2)+2\vvtrig\vvtwo\cos(4\dphi_2)\right]+\nonumber\\
&&Y_2(\dphi_2)\times B_1\left[1+2\vtrig\vone\cos(2\dphi_1)+2\vvtrig\vvone\cos(4\dphi_1)\right]-\nonumber\\
&&\hspace{-0.8in}B_1^2\left(
\ba{ll}
1&+2\vtrig\vone\cos(2\dphi_1)+2\vtrig\vtwo\cos(2\dphi_2)\\
 &+2\vvtrig\vvone\cos(4\dphi_1)+2\vvtrig\vvtwo\cos(4\dphi_2)\\
 &-2\vtrig\vone\vvtwo\cos2(\dphi_1-2\dphi_2)-2\vtrig\vtwo\vvone\cos2(2\dphi_1-\dphi_2)\\
 &-2\vone\vtwo\vvtrig\cos2(\dphi_1+\dphi_2)\\
 &+8(\vtrig)^2\vone\vtwo\cos(2\dphi_1)\cos(2\dphi_2)+8\vtrig\vvtrig\vone\vvtwo\cos(2\dphi_1)\cos(4\dphi_2)\\
 &+8(\vvtrig)^2\vvone\vvtwo\cos(4\dphi_1)\cos(4\dphi_2)+8\vtrig\vvtrig\vtwo\vvone\cos(4\dphi_1)\cos(2\dphi_2)
\ea
\right)\nonumber\\
&=&B_1[Y_2(\dphi_1)+Y_2(\dphi_2)-B_1]+\nonumber\\
&&B_1[Y_2(\dphi_1)-B_1]2\vtrig\vtwo\cos(2\dphi_2)+B_1[Y_2(\dphi_2)-B_1]2\vtrig\vone\cos(2\dphi_1)+\nonumber\\
&&B_1[Y_2(\dphi_1)-B_1]2\vvtrig\vvtwo\cos(4\dphi_2)+B_1[Y_2(\dphi_2)-B_1]2\vvtrig\vvone\cos(4\dphi_1)+\nonumber\\
&&\hspace{-0.8in}B_1^2\left(
\ba{l}
2\vtrig\vone\vvtwo\cos2(\dphi_1-2\dphi_2)+2\vtrig\vtwo\vvone\cos2(2\dphi_1-\dphi_2)+\\
2\vone\vtwo\vvtrig\cos2(\dphi_1+\dphi_2)-\\
8(\vtrig)^2\vone\vtwo\cos(2\dphi_1)\cos(2\dphi_2)-8\vtrig\vvtrig\vone\vvtwo\cos(2\dphi_1)\cos(4\dphi_2)-\\
8(\vvtrig)^2\vvone\vvtwo\cos(4\dphi_1)\cos(4\dphi_2)-8\vtrig\vvtrig\vtwo\vvone\cos(4\dphi_1)\cos(2\dphi_2)
\ea
\right).\label{eq:J2B2B3b}
\eea
As can be seen, the leading term in the background is the first term (first line of Eq.~(\ref{eq:J2B2B3b}) r.h.s.), of the order $B_1^2$. The terms on the second line of Eq.~(\ref{eq:J2B2B3b}) r.h.s. are of the order $\Jhat_2B_1v_2^2$. Since $\Jhat_2$ is of the order a couple of percent of $B_1$~\cite{uleryQM05,uleryHP} and $v_2^2$ is of the same order~\cite{flow4P}, those terms are of the order $B_1^2v_2^4$, or $\Jhat_2^2\sim\Jhat_3$. The terms on the third line are of the order $B_1^2v_2^6$ because $v_4\sim v_2^2$~\cite{flowPRC}, and those on the fourth line are a mixer of $B_1^2v_2^4$, $B_1^2v_2^6$, and smaller terms. Neglecting the terms of the order $B_1^2v_2^6$ or smaller, we obtain
\bea
\Jhat_2\otimes B_2+B_3
&=&B_1[Y_2(\dphi_1)+Y_2(\dphi_2)-B_1]+\nonumber\\
&&2B_1[Y_2(\dphi_1)-B_1]\vtrig\vtwo\cos(2\dphi_2)+2B_1[Y_2(\dphi_2)-B_1]\vtrig\vone\cos(2\dphi_1)+\nonumber\\
&&2B_1^2\vtrig\vone\vvtwo\cos2(\dphi_1-2\dphi_2)+2B_1^2\vtrig\vtwo\vvone\cos2(2\dphi_1-\dphi_2)+\nonumber\\
&&2B_1^2\vvtrig\vone\vtwo\cos2(\dphi_1+\dphi_2)-8B_1^2(\vtrig)^2\vone\vtwo\cos(2\dphi_1)\cos(2\dphi_2).\label{eq:J2B2B3c}
\eea
Except the first, leading term, all other terms in Eq.~(\ref{eq:J2B2B3c}) are of the same order of the three-particle correlation strength, $\Jhat_3$. Therefore, uncertainties on $v_2$ will result in uncertainty on the three-particle strength of the similar magnitude. 

In the above illustration, we have taken $B_1$ to be constant, not affected by the uncertainties on $v_2$. However, with the background normalization scheme of ZYAM or ZYA1, $B_1$ will be affected by the value of $v_2$~\cite{jetspec}. This effect on $B_1$, being obtained by normalization fixed at $\dphi=1$, works in the opposite direction as the $v_2$ uncertainty effect on the three-particle correlation strength, thereby making the effect of $v_2$ uncertainty smaller than illustrated above.

Experimentally, anisotropic flow is measured within finite centrality bins. The flow magnitudes vary with event multiplicity within a given finite centrality bin. Since flow correction is not linear in flow (for both two-particle and three-particle correlations), the calculated flow correction using the measured average magnitude of flow in a given centrality bin deviates from the true flow effect. The magnitude of this effect can be estimated by parameterizing the measured flow as a function of event multiplicity. With the measured multiplicity dependence of flow, the effect is very small and can be neglected~\cite{thesis}.

The major uncertainties on flow measurements are due to non-flow~\cite{flow4P}. The tranditional reaction plane measurements have large contributions from non-flow, so do the two-particle cumulant measurements. The non-flow effect is significantly reduced in the four-particle cumulant measurements because the jet multiplicity in our relavent jet energy range is small, mostly fewer than four particles. On the other hand, flow fluctuations reduce the measured four-particle cumulant $v_2$ and increase the reaction plane and two-particle cumulant $v_2$~\cite{raimond}. Due to the presence of other jets not correlated with the di-jet selected by the trigger particle, non-flow effect should be partially included in the flow correction in jet-correlation measurements. So the $v_2$ used in the correlation background should include the flow fluctuation effect but should not include the full non-flow effect. The interlink between flow and jet-correlation is complex; in the lack of any firmer solution, the average of the four-particle cumulant $v_2$ and the reaction plane $v_2$ has been used for background correction in jet-correlation measurements~\cite{jetspec}. The systematic uncertainty boundaries are set by the four-particle cumulant $v_2$ and the two-particle cumulant $v_2$. Here the two-particle, not the reaction plane $v_2$ is used as the systematic limit because the flow fluctuation (which is present in two-particle $v_2$) should be included in the background correction. However, in practice these two $v_2$ measurements are approximately equal.

\subsection{Systematics due to jet anisotropy\label{sec:jetAnisotropy}}

The hard-soft background, given by Eq.~(\ref{eq:HS}) or (\ref{eq:J2B2}), arises from the combinatorial pairing of a jet-correlated particle and a background particle. The background is modulated by anisotropic flow because the trigger particle and the background particles are all correlated to the reaction plane. In Eqs.~(\ref{eq:HS}) and (\ref{eq:J2B2}), the measured trigger $v_2$ is used. However, the anisotropy of the trigger particle, with the additional requirement that a jet-correlated associated particle present within the corresponding kinematic region, may not be equal to the measured anisotropy for inclusive trigger particles a priori. We will refer to those trigger particles with an associated partner as jet trigger particles, and refer to this effect as jet anisotropy. The other trigger particles (i.e., having no associated particles in the corresponding kinematic region) of course need to have anisotropy also different from the measured inclusive one, in order to have the inclusive one (including all trigger particles) as measured. 

Suppose the elliptic flow of the trigger particle accompanied by at least one associated particle is $\vjet$, and the fraction of those trigger particles is $\alpha$. Suppose the elliptic flow of all other trigger particles is $\vnjet$. The measured elliptic flow of inclusive trigger particles, $\vtrig$, requires:
\be\alpha\vjet+(1-\alpha)\vnjet=\vtrig.\label{eq:v2jv2o}\ee
The two-particle correlation normalized by the total number of trigger particles, given in Eq.~(\ref{eq:J2hatBg}) or (\ref{eq:J2hat}), is unchanged. The two-particle correlation for jet trigger particles, normalized per jet trigger, is however a factor of $1/\alpha$ of that given by Eq.~(\ref{eq:J2hatBg}) or (\ref{eq:J2hat}). The two-particle correlation for the other trigger particles, of course, vanishes. Only the jet trigger particles contribute to the hard-soft background; the other trigger particles do not contribute because of the vanishing associated particle multiplicity. The hard-soft background is thus given by
\be
\Jhat_2\otimes B_2=\alpha\left[
\ba{ll}
\frac{1}{\alpha}&\left(Y_2(\dphi_1)-B_1\left[1+2\vtrig\vone\cos(2\dphi_1)+2\vvtrig\vvone\cos(4\dphi_1)\right]\right)\times\nonumber\\
&\left(B_1\left[1+2\vjet\vtwo\cos(2\dphi_2)+2\vvjet\vvtwo\cos(4\dphi_2)\right]\right)+\nonumber\\
\frac{1}{\alpha}&\left(Y_2(\dphi_2)-B_1\left[1+2\vtrig\vtwo\cos(2\dphi_2)+2\vvtrig\vvtwo\cos(4\dphi_2)\right]\right)\times\nonumber\\
&\left(B_1\left[1+2\vjet\vone\cos(2\dphi_1)+2\vvjet\vvone\cos(4\dphi_1)\right]\right)
\ea
\right]
\label{eq:HSjet}
\ee
where the normalization is, as always, by the total number of trigger particles. The $1/\alpha$ factor in the two-particle correlation strength per jet trigger particle is cancelled by the fraction of the jet trigger particles of $\alpha$. Note that the $\vtrig$ ($\vvtrig$) in the background modulation (that is folded with the associated particle signal) in Eq.~(\ref{eq:HS}) and (\ref{eq:J2B2}) are replaced by $\vjet$ ($\vvjet$) in Eq.~(\ref{eq:HSjet}).

The soft-soft background, as given in Eqs.~(\ref{eq:B3new}), (\ref{eq:B3inctf}) and (\ref{eq:B3incflow}), is unchanged, because the $\vtrig$ in Eq.~(\ref{eq:B3inctf}) is replaced by $\alpha\vjet+(1-\alpha)\vnjet$ which is identical to $\vtrig$ by Eq.~(\ref{eq:v2jv2o}). 

Thus, the only change in the three-particle correlation analysis, due to the possible effect of different anisotropies of trigger particles with and without accompanying associated particles, is the hard-soft background term. The estimation of $\vjet$ is somewhat arbitrary. One may use a number of inputs to make an educated guess; these inputs include (i) two-particle correlation with respect to reaction plane, (ii) how large and small the elliptic flow of other trigger particles can have with considerations of hydrodynamic calculations, and (iii) how small the factor $\alpha$ can get without running into unreasonably large two-particle correlation strength per jet trigger particle. Another reasonable approach is to use, as an estimation of $\vjet$, the measured elliptic flow of particles at $\pt$ equal to the sum of those of the trigger particle and the accompanying associated particle.

However, there are further complications. We know that the flow background varies with the trigger particle azimuthal angle relative to the reaction plane~\cite{bielcikova}. Only when the two-particle jet-correlation is constant, is the hard-soft background from folding as in Eqs.(\ref{eq:HS}) and (\ref{eq:J2B2}) valid. In other words, Eqs.(\ref{eq:HS}) and (\ref{eq:J2B2}) have the implicit assumption of constant jet-correlation over the trigger particle orientation. Eq.(\ref{eq:HSjet}) has loosened this requirement in the sense that the jet-correlation magnitude can change with the trigger particle orientation (but still regularly according to harmonic osillations) through the different flow harmonics, $\vjet\neq\vtrig$. In real data, however, the jet-correlation may change with the trigger particle orientation in both its magnitude and shape. Such a change cannot be simply described by a differing $\vjet$ value. The hard-soft background can only be constructed from folding of the jet-correlation signal and background at each trigger particle orientation relative to the reaction plane separately and then integrated together over all trigger orientations. This requires measurements of two-particle jet-correlation as a function of trigger particle azimuthal angle relative to the reaction plane~\cite{kirill}.

\subsection{Induced Correlation between Associated and Background Multiplicities\label{sec:inducedCorrel}}

The hard-soft background term is obtained by folding the jet signal ($\Jhat_2$) with the background ($B_2$), both averaged over a given class (centrality) of events. The jet signal and the background level are not intrinsically correlated, but their magnitudes could be correlated via the event multiplicity. For instance, within a given centrality of events, if the jet signal increases with the event multiplicity (which in high multiplicity events is essentially as same as the background multiplicity), then the average of the product of the jet signal and the background level will not equal to the product of the individual averages of the jet signal and background level, i.e., $\mean{\Jhat_2B_2}\neq\mean{\Jhat_2}\mean{B_2}$. We shall estimate how large the difference, $\delta=\frac{\mean{\Jhat_2B_2}-\mean{\Jhat_2}\mean{B_2}}{\mean{\Jhat_2}\mean{B_2}}$, could be in central heavy-ion collisions by using simple assumptions of jet multiplicity as a function of event multiplicity, $N$, within a centrality bin $N_1<N<N_2$.

If the jet multiplicity linearly increases with $N^{1/3}$, $\Jhat_2=c_1+c_2(N/N_1)^{1/3}$, and since central heavy-ion event multiplicity is large, $B_2\approx N/2\pi-\Jhat_2\approx N/2\pi$, the difference can be expressed as $\delta=\frac{\mean{N^{4/3}}-\mean{N}\mean{N^{1/3}}}{\mean{N}\left(cN_1^{1/3}+\mean{N^{1/3}}\right)}$ where $c=c_1/c_2$. Obviously if $c_2=0$, i.e., there is no apparent correlation between $\Jhat_2$ and $B_2$, then $\delta$ is zero. If we assume constant cross-section over $N$ within $N_1<N<N_2$ so we can easily calculate the various averages, then we have $\delta=\frac{(6/7)\left(r^{7/3}-1\right)/(r+1)-(3/4)\left(r^{4/3}-1\right)}{(r-1)c+(3/4)\left(r^{4/3}-1\right)}$ where $r=N_2/N_1$. Let us plug in some reasonable numbers to get a feeling of how large the effect could be. If we take $N_1=400$ and $N_2=500$, then we get $\delta=\frac{0.000357}{0.25c+0.260}$. If $c_1=0$, then the relative difference is $\delta\approx 0.14\%$. Given the two-particle correlation signal to background ratio hence the three-particle signal to hard-soft background ratio of 1/30, this difference is about 5\% of the three-particle correlation signal. If $c_1=c_2$, then the difference is about a factor of 2 smaller.

If the jet signal linearly increases with the event multiplicity, $\Jhat_2=c_1+c_2N/N_1$, and again $B_2\approx N/2\pi-\Jhat_2\approx N/2\pi$, then the effect is larger because the apparent correlation between $\Jhat_2$ and $B_2$ is stronger. The relative difference can be obtained as $\delta=\frac{\mean{N^2}-\mean{N}^2}{\mean{N}\left(cN_1+\mean{N}\right)}$. Again for $c_2=0$, the difference is zero. Again assuming constant cross-section over $N_1<N<N_2$, we have $\delta=\frac{1-2r+r^2}{3(1+r)\left(2c+1+r\right)}$. Take the same example, $N_1=400$ and $N_2=500$, we have $\delta\approx\frac{0.00926}{2c+2.25}$. For $c_1=0$, the relative difference is $\delta\approx 0.4\%$, which amounts to 12\% of the three-particle correlation signal. If $c_1=c_2$, then again the effect is about a factor of 2 smaller.

We should note that the above estimates are on the conservative side, because the away-side associated particle multiplicity has a weak dependence on the event multiplicity~\cite{jetspec}. The dependence varies with the associated $\pt$. At low $\pt$ the associated particle multiplicity increases with the event multiplicity; at high $\pt$ the associated particle multiplicity decreases; at intermediate $\pt$ the dependence can be very weak.

For $pp$ and d+Au collisions, due to the significant trigger bias, the correlation between the two-particle correlation strength and the background multiplicity via event multiplicity can be large. Our estimate indicates that the effect can be as large as $\delta\sim 10\%$~\cite{thesis}. However, the hard-soft backgrounds in $pp$, d+Au, and peripheral Au+Au collisions are small, comparable to the three-particle correlation strength. So the effect is of the order of 10\% of the three-particle correlation strength. One may reduce the effect by dividing the $pp$ and d+Au events into several multiplicity bins.

\subsection{Centrality Bias of Triggered Events from Inclusive Distribution}

The underlying background in the triggered events may be different from the inclusive events because the triggered events may be biased towards larger multiplicity due to the requirement of a high $\pt$ trigger particle. In other words, the multiplicity distribution of the underlying background differs from that of the inclusive events. This effect can be studied by first obtaining the soft-soft background from the inclusive events as a function of the event multiplicity in single increment, and then weighting those soft-soft backgrounds with the multiplicity distribution of the triggered events. The multiplicity distribution of the triggered events should be offset towards the lower end by the amount corresponding to the jet associated multiplicity; this offset can be done in various ways to assess the systematic effect. 

Our estimate indicates that the systematic effect due to the slight mismatch in the multiplicity distributions in central Au+Au collisions is small and generally negligible~\cite{thesis}. The mismatch in the multiplicity distributions in the triggered events and inclusive (or minimum bias) events for $pp$, d+Au and very peripheral Au+Au collisions can be large. However, since the soft-soft background in these collisions is substantially smaller, the systematic effects due to the multiplicity distribution mismatch are also not large~\cite{thesis}. Nevertheless, one may subdivide the minimum bias events from $pp$ and d+Au collisions in several multiplicity classes, to the extent that is statistically meaningful, to reduce this effect.

Due to the different shapes of the multiplicity distributions, the flow correction obtained from inclusive events of a given centrality may also be slightly off. The anisotropic flow is measured in the predefined centrality windows, not for each event multiplicity value. Nevertheless, one may parameterize the measured flow as a function of the event multiplicity and assess the size of the effect it may cause. We estimate this effect to be very small and negligible~\cite{thesis}.

\subsection{Background Normalization\label{sec:systError.5}}

In two-particle correlation analyses, one often normalizes the background to the minimum of the raw signal so that the signal at minimum is zero (ZYAM). The STAR collaboration normalizes the background to the raw signal in the fixed range of $|\dphi\pm1|<0.2$, about where the signal is minimal (ZYA1)~\cite{jetspec}. The use of the fixed region is to avoid an additional running systematics, which for example could be affected by statistical precision of the data. One should note, however, this scheme of background normalization is only a working assumption, in lack of something better. Although the jet signal is expected to be non-negative everywhere in azimuth, the jet signal can be larger than zero in all azimuth if the jet peaks are broad and overlap, as likely the case in central Au+Au collisions.

On the other hand, three-particle correlation with jet signal in the two-dimensional $(\dphi_1,\dphi_2)$ angular space can provide a more stringent constraint on the background level, because the three-particle jet-like correlation should, in principle, have non-negative values anywhere on the two-dimensional $(\dphi_1,\dphi_2)$ plane. For example, when the three-particle correlation have some region where the signal vanishes, the projection onto one angular variable can in principle be above zero everywhere. In this case the background normalization to the minimum signal in the two-particle correlation over-subtracts the background, which could result in negative regions in the three-particle correlation function. 

\begin{figure*}[hbt]
\centerline{\psfig{file=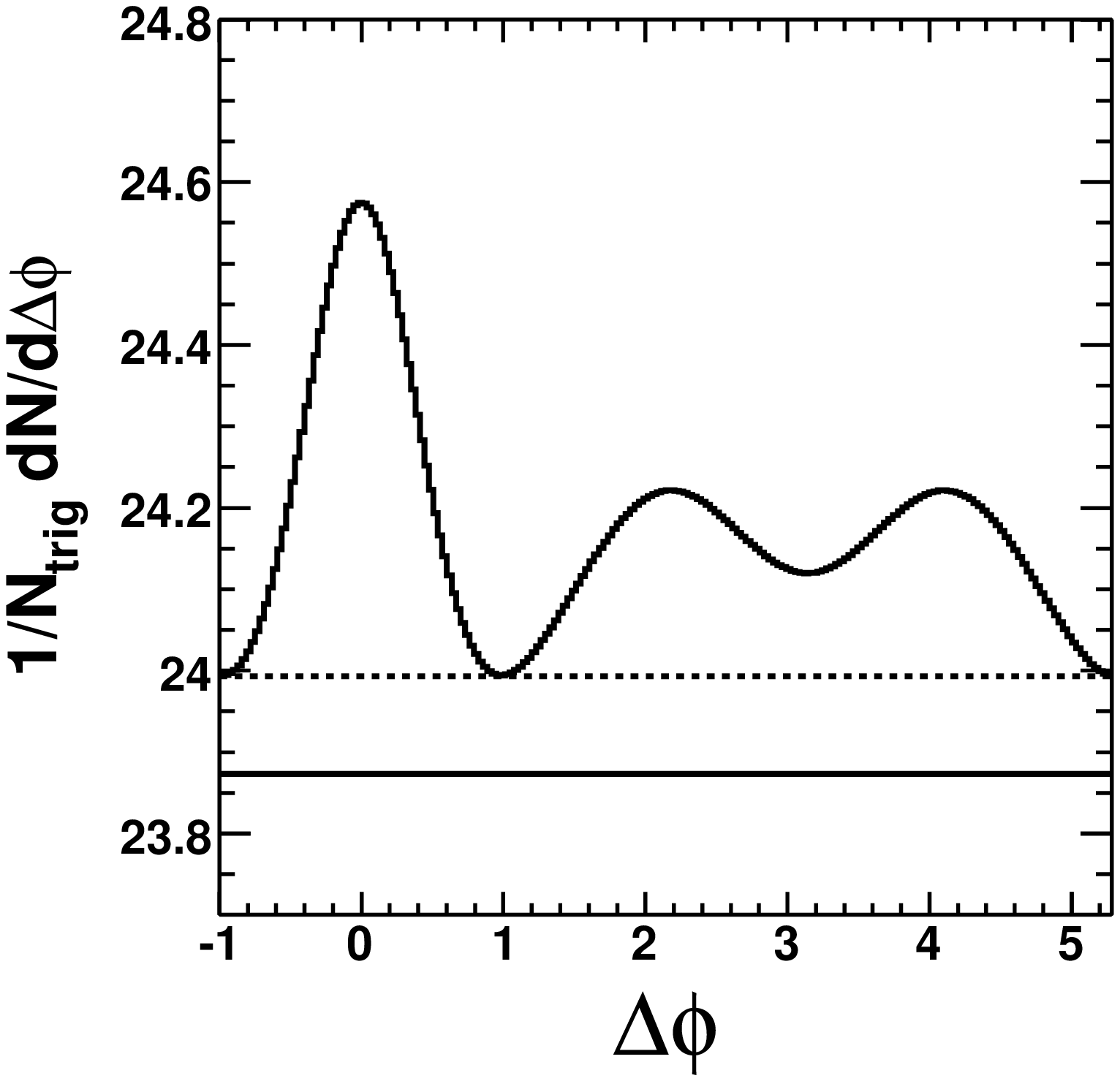,width=0.3\textwidth}
\psfig{file=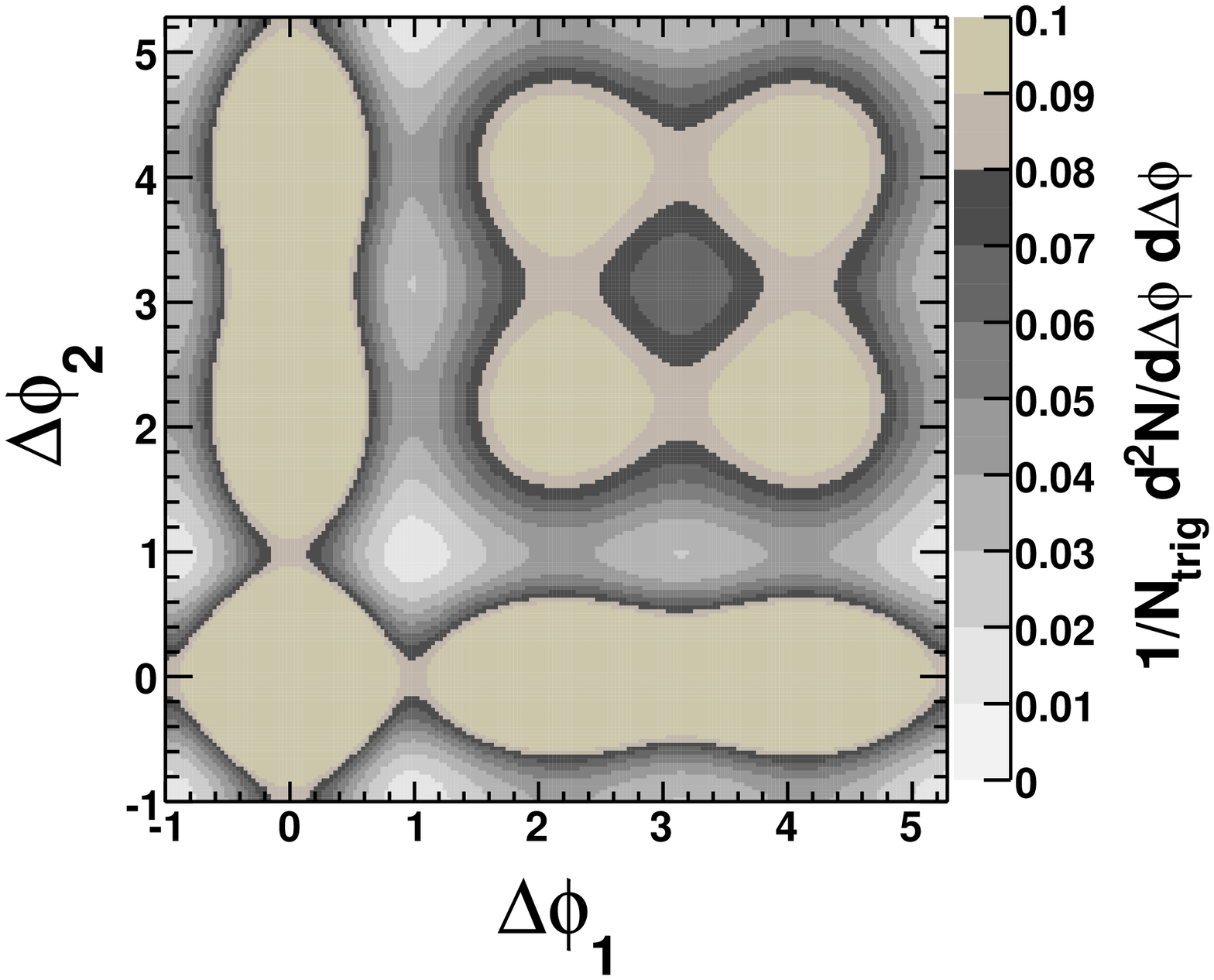,width=0.3\textwidth}
\psfig{file=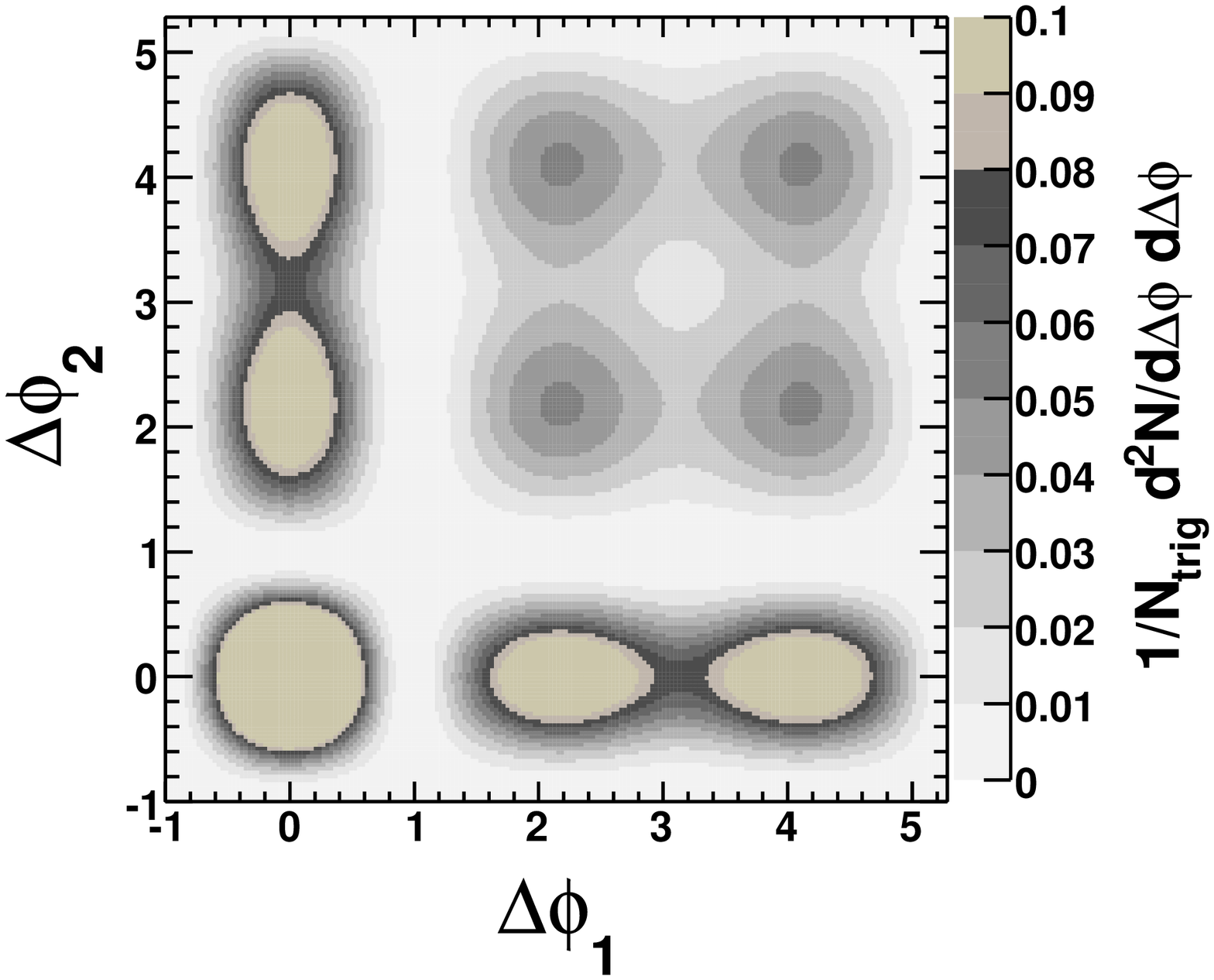,width=0.3\textwidth}
}
\caption{Left panel: two-particle correlation signal sitting on a flat background (solid horizontal line). The dashed horizontal line is the overestimated background level to match the signal at the minimum at $\dphi\approx 1$. Center panel: three-particle correlation with the true background subtracted (corresponding to solid horizontal line in the left panel). Right panel: three-particle correlation with two-particle background normalized to the signal at $\dphi\approx 1$ (corresponding to the dashed horizontal line in the left panel). The maxima of the three-particle correlation plots in the center and right panels are set to the same value 0.1 in order to compare the details of the minimum regions. The three-particle correlation functions themselves have larger maxima, about 0.50 and 0.35 for the center and right panels, respectively.}
\label{fig1}
\end{figure*}

Below we use a specific example to illustrate the above points. We define a jet-like two-particle correlation with a near-side peak and a broad double-hump away-side distribution:
\be
\Jhat_2(\dphi)=\frac{N_1}{\sqrt{2\pi}\sigma_1}\exp\left[-\frac{(\dphi)^2}{2\sigma_1^2}\right]+\frac{N_2/2}{\sqrt{2\pi}\sigma_2}\left(\exp\left[-\frac{(\dphi+\theta)^2}{2\sigma_2^2}\right]+\exp\left[-\frac{(\dphi-\theta)^2}{2\sigma_2^2}\right]\right).\label{eq:J2gaus}
\ee
We use $\sigma_1=0.4$, $\sigma_2=0.7$, $N_1=0.7$, $N_2=1.2$, and $\theta=1$. Obviously the correlation is positive everywhere. The minimum signal strength is at $\dphi\approx 1$ and is about $0.12$. It is not zero because of the overlap of the Gaussians. We suppose this jet-like two-particle correlation is presented in every event, and the background is flat in azimuthal angle at $B_0=dN/d\dphi=150/2\pi$. The raw two-particle correlation is shown in the left panel of Fig.~\ref{fig1}. The solid horizontal line shows the real background level, and the dashed horizontal line shows the normalized background level such that the jet-correlation signal is zero at $\dphi\approx 1$. Because of the overlap of the Gaussians, the normalized background is not the true background level but is overestimated. 

For illustration, we define the genuine three-particle correlation function to be the simple product of the two two-particle jet-correlation functions (i.e., the three-particle correlation function is factorized):
\be\Jhat_3(\dphi_1,\dphi_2)=\Jhat_2(\dphi_1)\Jhat_2(\dphi_2).\label{eq:Jhat3}\ee
This is shown in the center panel of Fig.~\ref{fig1}. In order to see the details of the minimum valley region, the plot maximum is set to 0.1; the three-particle correlation function itself has a larger maximum, about 0.50, which is on the near-side. Although in this specific, simple-minded example, the three-particle correlation cannot be zero if the two-particle correlation is non-zero, there are regions in the three-particle correlation function where the signal is almost zero as can been seen in the center panel of Fig.~\ref{fig1}. The two-particle correlation function, however, is significantly positive everywhere because one dimension is integrated out. Therefore, three-particle correlation is more powerful to determine the background than two-particle correlation. In principle this can be utilized to better normalize the background in two-particle jet-correlation. 

The three-particle signal after subtraction of the normalized (hence overestimated) background level is shown in the right panel of Fig.~\ref{fig1}. Again, in order to see the details of the minimum valley region, the plot maximum is set to 0.1; the three-particle correlation function itself has a larger maximum, about 0.35, which is on the near-side. As seen from the figure, the zero signal regions are now larger. For three-particle correlations more complex than the simple example here, the three-particle correlation function can have negative regions when the background normalization is done with ZYA1 or ZYAM for two-particle correlation. Again this is simply because one dimension of three-particle correlation is integrated out in two-particle correlation function. 

On the other hand, one may use three-particle ZYAM to obtain the background magnitude, i.e., the factor $a$, as aforementioned. This is a better estimation of the background because three-particle correlation contains richer, more differential information than two-particle correlation. The systematic uncertainty on $a$ may be indicated by the change in the value of $a$ from two-particle ZYAM to three-particle ZYAM. 

As discussed in Section~\ref{sec:realData.C}, the value of $b$ is determined from the associated multiplicity distribution in triggered events. The systematic uncertainty on $b$ may be assessed by varying the size of the azimuthal regions used for three-particle ZYAM normalization.

\section{Summary\label{sec:summary}}

In this paper, we have described an analysis method of jet-like three-particle azimuthal correlation between a high $\pt$ trigger particle and two softer particles. The method is devised with the particular two-component model in mind, in which the triggered event is made up by a di-jet and a bulk medium. The three-particle correlation analysis is extremely challenging due to the large combinatorial backgrounds in central relativistic heavy-ion collisions. Consequently, our emphasis is on the careful construction of the various backgrounds in an envisioned real data analysis and the evaluation of their major systematic uncertainties.  By doing so, we demonstrate that three-particle jet-correlation result from such an analysis method is rather robust even though the background is large. The key to achieve the robustness is to use real data to construct the background terms self-consistently with only two well-defined scaling factors: the factor $a$ to scale the inclusive event multiplicity to the background multiplicity underlying the triggered events by the ZYA1 or ZYAM assumption, and the factor $b$ to take into account any difference between the multiplicity distributions of the inclusive events and the underlying background events such that the final three-particle correlation has zero yield at minimum.

\section*{Acknowledgments}

We thank our STAR collaborators, in particular, Dr.~Carl Gagliardi, Dr.~Bedanga Mohanty, Dr.~Claude Pruneau, Dr.~Marco van Leeuwen, Dr.~Sergei Voloshin, and Dr.~Zhangbu Xu for valuable discussions. We thank Dr.~Andrew Hirsch for reading of the manuscript. This work is supported by U.S. Department of Energy under Grants DE-FG02-02ER41219 and DE-FG02-88ER40412.



\begin{thebibliography}{99}
\bibitem{noteRP}
In principle, one can take care of reaction plane orientation in event-mixing, by mixing events with similar measured reaction plane angles, or by rotating all particles in the event so that all events have the same measured reaction plane angle, and then correct for the measured reaction plane resolution in the final result. However, the correction for the reaction plane resolution is non-trivial.
\end{thebibliography}
\end{document}